\documentclass[journal]{IEEEtran}
\def\BibTeX{{\rm B\kern-.05em{\sc i\kern-.025em b}\kern-.08em
    T\kern-.1667em\lower.7ex\hbox{E}\kern-.125emX}}
\usepackage{amsmath}
\usepackage{amsthm}
\usepackage{amsfonts}
\usepackage{amssymb}
\usepackage{graphicx}
\usepackage{cite}
\usepackage{subfigure}
\usepackage{balance}
\usepackage{algorithm}
\usepackage{algorithmic}
\usepackage{enumerate}
\usepackage{color}
\usepackage{array}
\usepackage{indentfirst}
\usepackage{multirow}
\usepackage{booktabs}
\usepackage{color}
\usepackage{slashbox}
\usepackage{diagbox}
\usepackage{threeparttable}
\usepackage[svgnames,table]{xcolor}
\usepackage{bm}
\usepackage{epstopdf}
\usepackage{booktabs}
\usepackage{multirow}

\graphicspath{{figures/}}

\makeatletter

\newcommand{\Rmnum}[1]{\expandafter\@slowromancap\romannumeral #1@}
\makeatother

\begin{document}
\title{Underwater Target Recognition based on Multi-Decision LOFAR Spectrum Enhancement: A Deep Learning Approach}

\author{\IEEEauthorblockN{Jie Chen, \emph{Member, IEEE}, Jie Liu, Chang Liu, \emph{Member, IEEE}, Jian Zhang, and Bing Han } 


\thanks{The authors are with the University of Electronic Science and Technology of China, Chengdu, China
}

%

}

\maketitle
\vspace{-1.6cm}

\begin{abstract}
The Low frequency analysis and recording (LOFAR) spectrum is one of the key features of the under water target, which can be used for underwater target recognition. However, the underwater environment noise is complicated and the signal-to-noise ratio of the underwater target is rather low, which introduces the breakpoints to the LOFAR spectrum and thus hinders the underwater target recognition. To overcome this issue and to further improve the recognition performance, we adopt a deep learning approach for underwater target recognition and propose a LOFAR spectrum enhancement (LSE)-based underwater target recognition scheme, which consists of preprocessing, offline training, and online testing. In preprocessing, a LOFAR spectrum enhancement based on multi-step decision algorithm is specifically designed to recover the breakpoints in LOFAR spectrum. In offline training, we then adopt the enhanced LOFAR spectrum as the input of convolutional neural network (CNN) and develop a LOFAR-based CNN (LOFAR-CNN) for online recognition. Taking advantage of the powerful capability of CNN in feature extraction, the proposed LOFAR-CNN can further improve the recognition accuracy. Finally, extensive simulation results demonstrate that the LOFAR-CNN network can achieve a recognition accuracy of $95.22\%$, which outperforms the state-of-the-art methods.

\end{abstract}


\section{Introduction\label{sect: intr}}
\IEEEPARstart{T}{he} vast ocean contains rich mineral resources, marine living resources and chemical resources which can be exploited for economic benefits.
Thus, the marine developments, e.g., the submarine prospecting, the oil platform monitoring, and the economic fish detection, are of great importance.
Specifically, one of the key tasks in marine developments is the underwater target recognition \cite{yang2019deep, jouhari2019underwater}.
Deep learning (DL)-based underwater target recognition is a new way of realizing underwater target recognition in addition to the existing recognition methods to extract features and train classifiers manually. By using this method, it can automatically extract features from the original signal, compress feature vectors, fit the target map, reduce the impact of noise,  avoid feature loss during manual extraction, improve generalization capabilities, and constantly improve the efficiency and accuracy of identification during the model process.

Recently, DL techniques \cite{goodfellow2016deep} have been exploited for wireless physical layer \cite{wang2017deep, liu2019deep, liu2020deeptransfer, lxm2020deepresidual} and many effective and efficient DL-based schemes have been proposed for underwater target recognition.
For example, \cite{jin2020deep} and \cite{liu2019expansion} focused on underwater target recognition which didn't have insufficient training samples. In the first step, the original audio was converted into LOFAR spectrum, and then generative adversarial networks (GAN) was used for sample expansion. In the second step, a $15\%$ performance improvement could be obtained by using convolutional neural networks (CNNs) for feature learning and classification when the number of samples was more sufficient. \cite{yang2018competitive} combined competitive learning with
deep belief network (DBN) and proposed a deep competitive network that used unlabeled samples to solve small number of samples in acoustic target recognition. This method could achieve a classification accuracy of $90.89\%$. To address the negative impact of redundant features on recognition accuracy and efficiency, the authors in \cite{shen2018compression}  proposed a compressed deep competition network which combined network pruning with training quantization and other technologies and could achieve a classification accuracy of $89.1\%$. \cite{yan2018resonance, ke2018underwater} proposed a new time-frequency feature extraction method by jointly exploiting the resonance-based sparse signal decomposition (RSSD) algorithm, the phase space reconstruction (PSR), the time-frequency distribution (TFD), and the manifold learning.
At the same time, a one-dimensional convolutional auto-encoder-decoder model was used to further extract and separate features from high resonance components, which finally completed the recognition task and achieves a recognition accuracy of $93.28\%$. In addition, \cite{zhu2017deep, mcquay2017deep, hu2018deep} all used convolutional neural networks for feature extraction, but the application scenarios and the classifiers were different. \cite{zhu2017deep} proposed an automatic target recognition method of unmanned underwater vehicle (UUV), which adopted CNN to extract features from sonar images and used support vector machine (SVM) classifier to complete the classification. \cite{mcquay2017deep} aimed to study different types of marine mammals. It also used the CNN+SVM structure to complete the feature extraction and classification recognition task. It compared the two classification and multi-class task scenarios. \cite{hu2018deep} adopted the civil ship data set and exploited the framework structure of CNN+ELM (extreme learning machine) as the underwater target classifier, which improved the recognition accuracy. We can see that with the in-depth research of scholars, the recognition rate of underwater targets based on deep learning has gradually increased.


\begin{figure*}[t]
	\centering
	\includegraphics[width=\linewidth]{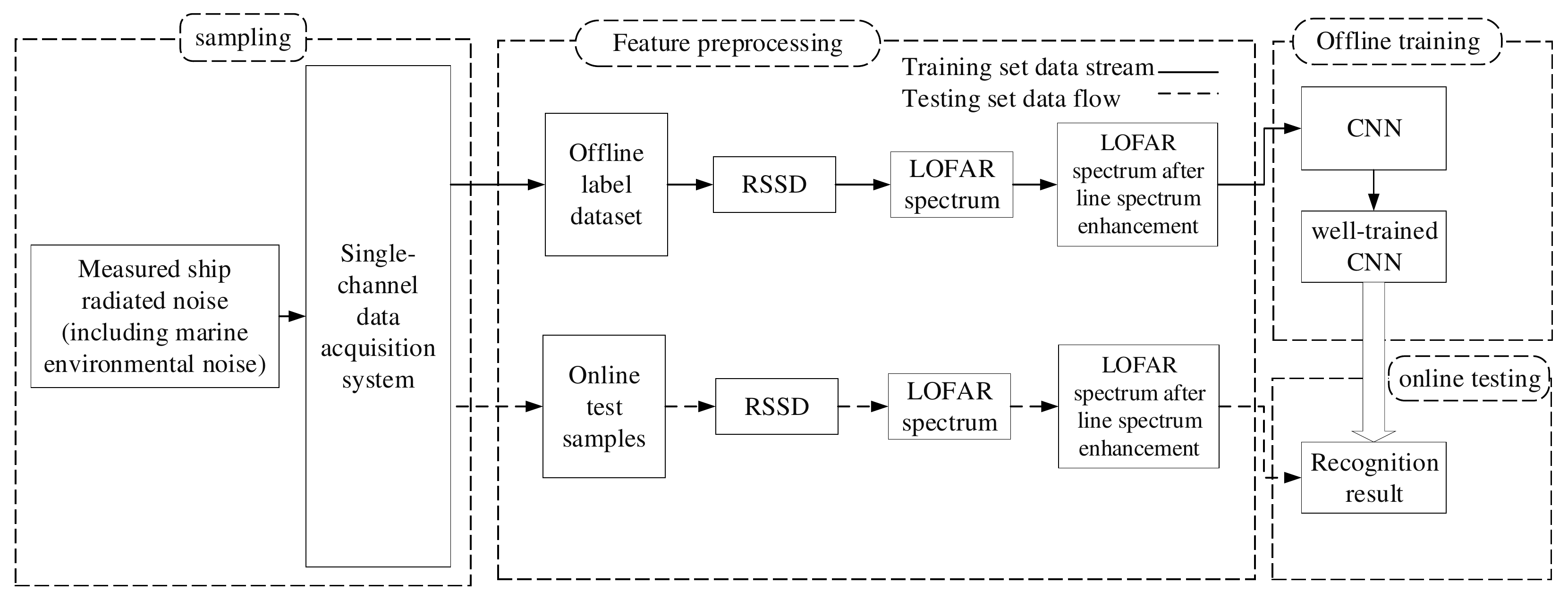}
	\caption{Deep learning underwater target recognition framework based on multi-step decision LOFAR line spectrum enhancement.}
	\label{Fig:framework}
\end{figure*}

Note that the collected raw data is always seriously polluted by environmental noise, which introduces the breakpoints to the LOFAR spectrum.
The so-called breakpoints directly affect the subsequent feature extraction and thus degrade the performance of the subsequent signal processing.
Motivated by this, we first adopt a model-based approach to recover the breakpoints in LOFAR  spectrum and then adopt a DL approach for task recognition.
The main contributions of this paper are as follows.
\begin{enumerate}[(1)]
\item Different from the traditional algorithm, we use the decomposition algorithm based on resonance signal to preprocess the signal. Based on the multi-step decision algorithm with the line spectrum characteristic cost function \cite{hubel1962receptive}, this paper proposes the specific calculation method of double threshold. In the purpose, this algorithm not only retains the continuous spectrum information in the original LOFAR spectrum, but also merges the extracted line spectrum with the original LOFAR spectrum. Finally, the breakpoint completion of the LOFAR spectrum is realized.

\item In order to further improve the recognition rate of underwater targets, we adopt the  enhanced  LOFAR  spectrum  as  the  input of CNN and develop a LOFAR-based CNN (LOFAR-CNN) for online recognition. Taking advantage of the powerful capability of CNN in feature extraction, the proposed LOFAR-CNN can further improve the recognition accuracy.

\item Simulation results demonstrate that when testing on the ShipsEar database \cite{santos2016shipsear},  our proposed LOFAR-CNN method can achieve a recognition accuracy of $95.22\%$ which outperforms the state-of-the-art methods.
\end{enumerate}

The rest of this article is organized as follows. The second section introduces the model of the system. The third section introduces the deep learning underwater target signal recognition framework based on multi-step decision LOFAR line spectrum enhancement. The fourth section is the experimental verification and simulation results of our proposed algorithm framework. The fifth section is the summary of the article.

Some notations in this paper are shown in the following. $\|\cdot\|_2$ and $\|\cdot\|_1$ respectively represent the L2 norm and L1 norm. $STFT\{\cdot\}$ is short-time Fourier transform. Term $E(\cdot)$ is the statistical expectation. $argmin$ represents the variable value when the objective function is minimized.


\section{SYSTEM MODEL}
In this paper, we consider a deep learning underwater target recognition framework based on multi-step decision LOFAR line spectrum enhancement which is shown in Fig.~\ref{Fig:framework}. It is divided into four modules: sampling, feature preprocessing, offline training and online testing.

\subsection{Signal Decomposition Algorithm based on Resonance}


In traditional signal processing, Fourier transform is usually used to analyze in the frequency domain or time-frequency domain, but these methods are only valid for periodic stationary signals. However, due to the generation mechanism of ship radiated noise and the complex channel conditions in the marine environment, the ship radiated noise collected by hydrophones is usually the mixture of oscillating signals and transient non-oscillating signals \cite{yan2018resonance}. The harmonic component (or oscillation component) of the ship's radiated noise plays an important role in the identification of underwater targets. Therefore, a signal decomposition algorithm based on resonance that effectively responds to nonlinear signals is used to preprocess the signal. Based on the oscillation characteristics rather than the frequency or scale, the method can obtain a signal composed of multiple simultaneous and continuous oscillations (high resonance component). To some extent, it weakens the transient non-oscillation signal of uncertain duration (low resonance component) and gaussian white noise (residual component) which is conducive to feature extraction.


The RSSD algorithm regards resonance as the basis for signal decomposition \cite{huang2017feature}, and the $Q$ factor quantifies the degree of signal resonance. Specifically, high-resonance signals exhibit a higher degree of frequency aggregation in the time domain, more simultaneous oscillating waveforms with a larger $Q$ factor. Low-resonance signals appear non-oscillating and indefinite transient signal with a smaller $Q$ factor. Therefore, the basic theory of the RSSD algorithm is that by using two different wavelet basis functions (corresponding to $Q$ factors of different sizes), we can find a sparse representation of a complex signal and reconstruct the signal.

\begin{figure}[t]
	\centering
	\includegraphics[width=0.86\linewidth]{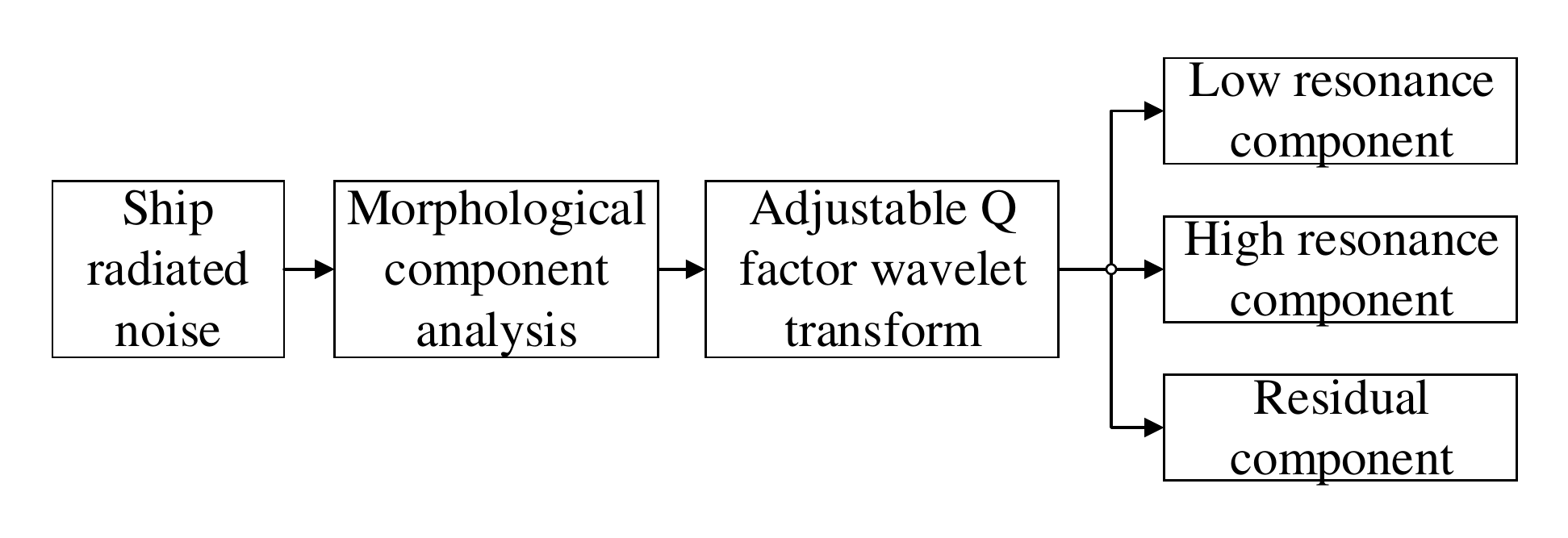}
	\caption{Signal decomposition algorithm based on resonance.}
	\label{Fig:Signal_decomposition_algorithm}
\end{figure}

The algorithm mentioned in this section is divided into adjustable Q-Factor Wavelet Transform (TQWT) \cite{selesnick2011wavelet} and Morphological Component Analysis (MCA) \cite{starck2005image}. Its algorithm framework is shown in Fig.~\ref{Fig:Signal_decomposition_algorithm}.

1)~\textbf{Morphological component analysis}


Morphological component analysis is usually used to decompose signals with different morphological characteristics \cite{al2009rolling}. The ship radiated noise with oscillating and non-oscillating component has different morphological characteristics. So the MCA algorithm can be used to separate and extract the ship radiated noise in order to construct the optimal sparse representation for its high resonance and low resonance component.

Considering the discrete ship radiated noise sequence, the signal can be sparsely expressed as:
\begin{equation}\label{}
x = {\Phi _h}{w_h} + {\Phi _l}{w_l} + n,
\end{equation}
where $w_h$, $w_l$ are the wavelet coefficients corresponding to the high resonant component $x_h$ and the low resonant component $x_l$. $\Phi _h$, $\Phi _l$ are wavelet basis functions corresponding to $x_h$, $x_l$. $n$ represents the residual components of the signal which removes first two.

The purpose of MCA is to obtain an optimal representation $w_h$, $w_l$ of the high-resonance component and low-resonance component of the signal. This problem can be solved by minimizing the following objective function:
\begin{equation}\label{}
\begin{split}
J({w_l},{w_h}) = & \| {x - {\Phi _h}{w_h} - {\Phi _l}{w_l}} \|_2^2 + \sum\limits_{j = 1}^{{J_h} + 1} {{\lambda _{h,j}}{{\| {w_h^j} \|}_1}}  \\
&+ \sum\limits_{j = 1}^{{J_l} + 1} {{\lambda _{l,j}}{{\| {w_l^j} \|}_1}}.
\end{split}
\end{equation}
Here, $J_h$ and $J_l$ represent the number of decomposition layers of $x_h$ and $x_l$. $w_h^j$ and $w_l^j$ are the wavelet coefficients of the high resonance component and the low resonance component of the $j$th layer, respectively. $\lambda _{h,j}$, $\lambda _{l,j}$ are the normalized coefficients of $w_{h,j}$, $w_{l,j}$ and their values are related to energy of $\Phi_{h,j}$, $\Phi_{l,j}$:
\begin{equation}\label{}
\lambda _{l,j} = k_{l,j} \| {\Phi _{l,j}} \|_2,j = 1,2, \cdots , J_l + 1,
\end{equation}
\begin{equation}\label{}
\lambda _{h,j} = k_{h,j} \| {\Phi _{h,j}} \|_2,j = 1,2, \cdots , J_h + 1,
\end{equation}
where $k_{l,j}$, $k_{h,j}$, $(k_{l,j}+k_{h,j} = 1)$ are the proportionality coefficient of the energy distribution of the high resonance component and the low resonance component. $k_{l,j} = k_{h,j}=0.5$ are selected to balance the energy distribution of the two components.

Through decomposition of the Augmented Lagrangian Shrinkage Algorithm (SALSA) \cite{huang2017feature}, the optimal wavelet coefficients can be obtained by solving the optimization problem of the formula. Therefore, the optimal expressions for the high resonance component and the low resonance component obtained by the MCA algorithm are:
\begin{equation}\label{}
\begin{array}{l}
x_h^* = \Phi _h^{}w_h^*,
\end{array}
\end{equation}
\begin{equation}\label{}
\begin{array}{l}
x_l^* = \Phi _l^{}w_l^*.
\end{array}
\end{equation}

In summary, the purpose of the RSSD algorithm is to construct the optimal sparse representation of the high and low resonance components of the ship radiated noise. The specific steps can be expressed as follows:

1) Select the appropriate filter scaling factor $\alpha$, $\beta$ according to the waveform characteristics of the signal. Then calculate the parameters $Q_h$, $r_h$, $J_h$ corresponding to the high resonance component, and the parameters $Q_l$, $r_l$, $J_l$ corresponding to the low resonance component. At last, construct the corresponding wavelet basis function $\Phi_h$, $\Phi_l$.



2) Reasonably set the weighting coefficient $\lambda_{h,j}$, $\lambda_{l,j}$ of the L1 norm of the wavelet coefficients of each layer. Obtain the optimal wavelet coefficient $w_h^*$, $w_l^*$ by minimizing the objective function through the SALSA algorithm.


3) Reconstruct the optimal sparse representation $x_h^*$, $x_l^*$ of high resonance components and low resonance components.

2)~\textbf{Adjustable $Q$ factor wavelet transform}

\begin{figure}[t]
	\centering
	\includegraphics[width=0.86\linewidth]{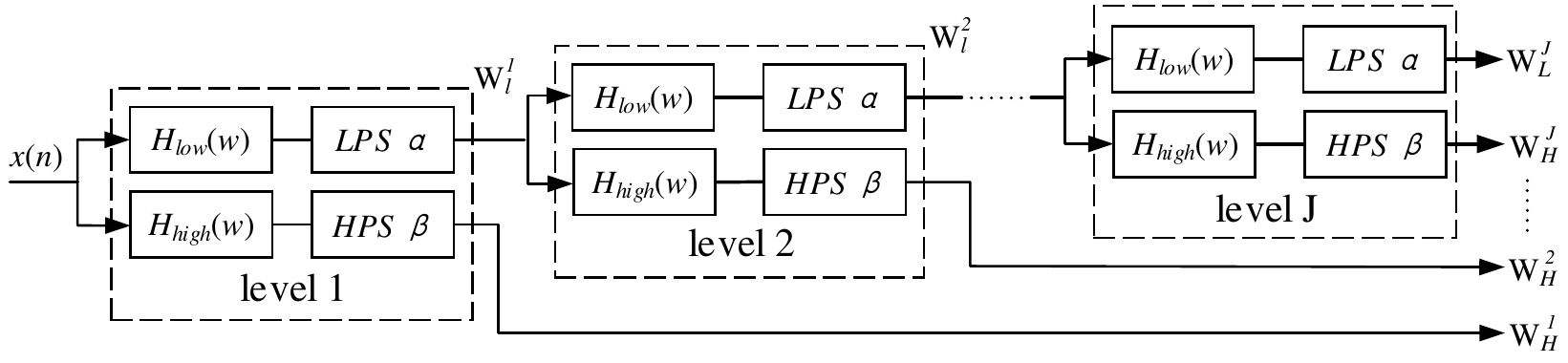}
	\caption{Analysis filter bank.}
	\label{Fig:3}
\end{figure}

\begin{figure}[t]
	\centering
	\includegraphics[width=0.86\linewidth]{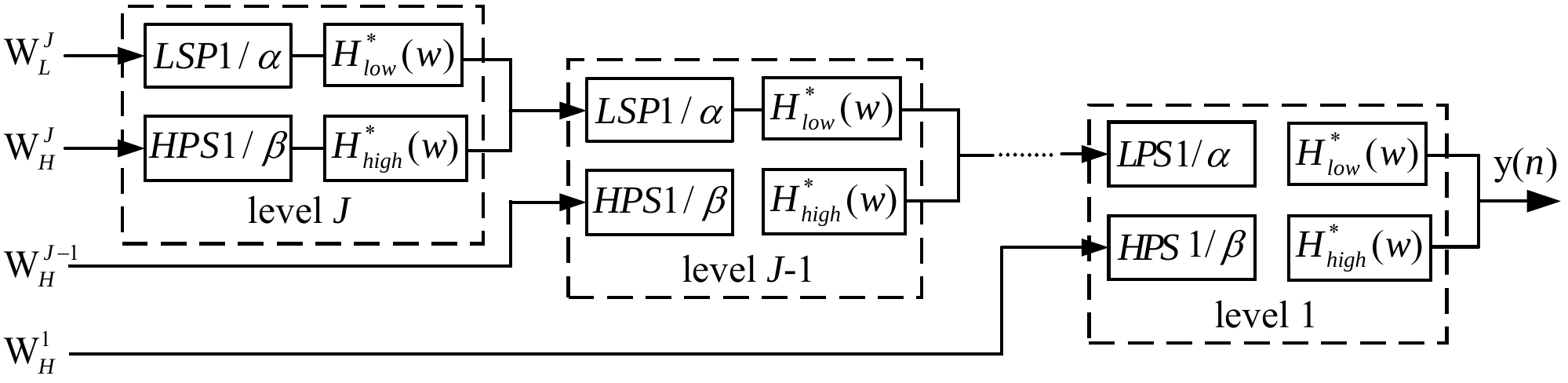}
	\caption{Integrated filter bank.}
	\label{Fig:4}
\end{figure}

TQWT is a discrete wavelet transform that can flexibly adjust the constant $Q$ factor according to the resonance of the processed signal, which has an overcomplete basis and can be perfectly reconstructed \cite{shensa1992discrete}. This section uses the TQWT toolbox to complete simulation experiments and signal processing. The implementation framework consists of two filter banks which are analysis filter bank and integrated filter bank. They are shown in Fig.~\ref{Fig:3} and Fig.~\ref{Fig:4}.

The analysis filter bank of each layer is composed of high-pass filter $H_{high}(w)$, low-pass filter $H_{low}(w)$, and the corresponding scaling process, which are defined as follows:
\begin{equation}\label{}
{H_{high}}(w)=\left\{\begin{array}{ll}
0 & \left |w\right |\leq (1-\beta )\pi    \\
\theta (\frac{\alpha \pi -w}{\alpha +\beta -1}) & (1-\beta)\pi\leq w\leq \alpha \pi\\
1 & \alpha \pi \leq \left | w \right |  \leq \pi
\end{array}\right.
,
\end{equation}
\begin{equation}\label{}
{H_{low}}(w)=\left\{\begin{array}{ll}
1 & \left |w\right |\leq (1-\beta )\pi    \\
\theta (\frac{w+ (\beta-1) \pi }{\alpha +\beta -1}) & (1-\beta)\pi\leq w\leq \alpha \pi\\
0 & \alpha \pi \leq \left | w \right |  \leq \pi
\end{array}\right.
.
\end{equation}
$\theta (w) = 0.5(1 + cos(w))\sqrt {2 - \cos (w)}$ is the Daubechies filter with second-order disappearing moment \cite{selesnick2011wavelet}.  $\alpha$, $\beta$ $(0 < \alpha  < 1, 0 < \beta  < 1)$ are the scaling factors after the signal passes through the low-pass and high-pass filters, respectively. The scaling process of low-pass and high-pass are defined as:
\begin{equation}\label{}
Y(w) = X(\alpha w), \left| w \right| \le \pi,
\end{equation}
\begin{equation}\label{}
Y(w) = \left\{ \begin{array}{ll}
X(\beta w + (1 - \beta )\pi ) & 0 \le w \le \pi\\
X(\beta w - (1 - \beta )\pi ) & -\pi < w < 0
\end{array} \right.
.
\end{equation}

The $Q$ factor quantifies the degree of signal resonance, and its definition is ${f_c}/BW$, where $f_c$ represents the center frequency of the signal and $BW$ represents the bandwidth.

If the sampling frequency of the original input signal is $f_s$, then the center frequency $f_c$, the filter bank level $j$ and $\alpha$, $\beta$ \cite{wang2014feature} can be expressed as:
\begin{equation}\label{}
{f_c} = {\alpha ^j}\frac{{2 - \beta }}{{4\alpha }}{f_s}
.
\end{equation}
Similarly, bandwidth $BW$ can be expressed as:
\begin{equation}\label{}
BW = 0.5\beta {\alpha ^{j - 1}}\pi
.
\end{equation}
Therefore, the $Q$ factor is derived as:
\begin{equation}\label{}
Q = \frac{{2 - \beta }}{\beta }
.
\end{equation}

After the original signal passes through the filter bank, the output of the low-pass channel is iteratively inputted to the deeper level filter bank until the preset level $J$. At the same time, the wavelet basis functions ${\Phi _h}$, ${\Phi _l}$ are constructed by selecting the oversampling rate $r$. The deepest level ${J_{\max }}$ and the oversampling rate $r$ are defined as follows:
\begin{equation}\label{}
r = \frac{\beta }{{\alpha + 1}}
,
\end{equation}
\begin{equation}\label{}
{J_{\max }} = \left| {\frac{{\log (\beta N/8)}}{{\log (1/\alpha )}}} \right|
.
\end{equation}

In summary, in the TQWT algorithm, $Q$, $r$, $J$ can be calculated by selecting $\alpha$, $\beta$, and $\alpha$, $\beta$ selection is only determined by the inherent oscillation characteristics of the signal. Therefore, it can flexibly select $\alpha$, $\beta$ according to the specific requirements of $Q$, $r$, $J$. For the input signal of ship radiated noise, we need to set $Q_h$, $r_h$, $J_h$ in order to extract its high resonance information and set $Q_l$, $r_l$, $J_l$ to extract its low resonance information.


\section{LOFAR SPECTRAL LINE ENHANCEMENT BASED ON MULTI-STEP DECISION}


The line spectrum has been widely used in the field of passive sonar ship target recognition because of its significant sound source information and relatively high signal-to-noise ratio. The Low Frequency Analysis Representation (LOFAR) spectrum transforms the signal received by the passive sonar from time domain to time-frequency domain by using the short-time Fourier transform (STFT), which can reflect the signal in the two dimensions of time domain and frequency domain. Scientists observe the line spectrum in the LOFAR spectrum to determine the presence or absence of the target, and perform tracking and recognition \cite{goodfellow2016deep}. Because there are more demand of the stealth technology of the ship and the radiated noise of the ship’s target is greatly reduced, the signal-to-noise ratio of the ship radiated noise received by the hydrophone array is also decreasing. The line spectrum components get more difficult to identify. There are a large number of research results on automatic detection and extraction of line spectrum under low signal-to-noise ratio.

In this paper, we study from the multi-step decision algorithm based on the line spectrum feature cost function proposed by Di Martino \cite{di1993lofargram}. Then we propose a specific calculation method of double threshold, and retain the continuous spectrum information in the original LOFAR spectrum. At last, we combine the original LOFAR spectrum with the extracted line spectrum, and complete the recognition and detection of underwater target by making full use of the advantages of deep neural network feature extraction.

\subsection{Structure LOFAR Spectrum}


The LOFAR spectrum is calculated by short-time Fourier transform (STFT). Unlike the traditional Fourier transform, which requires signal stability, STFT is suitable for non-stationary signals. It takes advantage of the short-term stationary characteristics of the signal. After windowing and framing the signal, the Fourier transform is performed to obtain the signal at time-frequency. Then it is more accurately characterize the distribution of signal frequency components and time nodes. The calculation formula is as follows:
\begin{equation}\label{}
STFT\{s(t)\}= \int_{ - \infty }^\infty  {s(t)w(t - \tau )} {e^{ - jwt}}dt,
\end{equation}
where $STFT\{\cdot\}$ is short-time Fourier transform, $s(t)$ is the signal to be transformed and $w(t)$ is the window function (truncating function). The specific calculation steps are as follows:

(1) Framing and windowing. Divide the sampling sequence of the signal into $K$ frames and each frame contains $N$ sampling points. Due to the correlation between the frames, there are usually some points overlap between the two frames. Framing is equivalent to truncating the signal, which will cause distortion of its spectrum and leakage of its spectral energy. In order to reduce spectral energy leakage, different truncation functions  which are called window function can be used to truncate the signal. The practical application window functions include Hamming window, rectangular window and Hanning window, etc.


(2) Normalization and decentralization. The signal of each frame needs to be normalized and decentralized, which can be calculated by the following formula:
\begin{equation}\label{}
s''(t) = \frac{s(t) - E[s(t)]}{\max(|{s'(t)|)}}.
\end{equation}
Here, $s^{'}(t)$ is the normalization of $s(t)$, which makes the power of the signal uniform in time. $s^{''}(t)$ is the decentralization of $s(t)$, which makes the mean of the samples zero.

(3) Perform Fourier transform on each frame signal and arrange the transformed spectrum in the time domain to obtain the LOFAR spectrum.

\subsection{Analysis and Construction of Line Spectrum Cost Function}

The definition of the line spectrum feature cost function is as follows:
\begin{equation}\label{}
O(\eta)= \frac{\lambda F(\eta) + \mu T(\eta)}{A(\eta)},
\end{equation}
where $\eta$ represents a summation path along the time axis in the observation window of the LOFAR graph, and the length of the path is $N$. $A(\eta)$ characterize the amplitude characteristics of the line spectrum, $F(\eta)$ is the frequency continuity of the line spectrum, and $T(\eta)$ is the trajectory continuity of the line spectrum, $\lambda$ and $\mu$ are weighting coefficients. The definitions of $A(\eta)$, $F(\eta)$, and $T(\eta)$ are as follows:
\begin{equation}\label{}
A(\eta ) = \sum\limits_{i = 1}^N {a({P_i})}
,
\end{equation}
\begin{equation}\label{}
F(\eta ) = \sum\limits_{i = 3}^N {\left| {d({P_{i - 2}},{P_{i - 1}}) - d({P_{i - 1}},{P_i})} \right|}
,
\end{equation}
\begin{equation}\label{}
T(\eta ) = \sum\limits_{i = 1}^N {g({P_i})}
.
\end{equation}
Each pixel on the summing path is ${P_i}(1 \le i \le N)$, which means a point on the $i$ line of the time axis. $a({P_i})$ characterizes the amplitude of the point $P_i$. $d({P_{i - 1}}, {P_i})$ characterizes the frequency gradient at two points in the path, which is defined as follows:
\begin{equation}\label{}
d({P_{i - 1}},{P_i}) = f({P_{i - 1}}) - f({P_i}),
\end{equation}
where $f({P_i})$ represents the frequency of the point $P_i$. $g(P_i)$ characterizes the breakpoint identification, which is defined as follows:
\begin{equation}\label{}
g({P_i}) = \left\{ \begin{array}{ll}
1 & a({P_i}) < \varepsilon \\
0 & others
\end{array} \right.
.
\end{equation}
If the amplitude of the point $P_i$ is less than $\varepsilon$, it is regarded as a breakpoint and recorded as $1$, otherwise it is recorded as $0$. Regarding the calculation of the threshold $\varepsilon$, the original algorithm is mostly set by empirical values, and a new calculation method is proposed as follows:
\begin{equation}\label{}
P(w) = {\left| {STFT\left\lbrace n(t)\right\rbrace  } \right|^2}
,
\end{equation}
\begin{equation}\label{}
\varepsilon  = A_{average} = \sqrt {\frac{{\sum {P(w)} }}{{M * N}}},
\end{equation}
where $n(t)$ represents the marine environmental noise. The sampling sequence of the interference noise in the marine environment is subjected to STFT transformation which can obtain the LOFAR spectrum. At the same time, the instantaneous power $p(w)$ of each time-frequency point is calculated. $M$, $N$ represent the points of frequency domain and time domain of LOFAR spectrum. The power of all time and frequency points is summed and averaged to obtain the average power. Take a square to get the average amplitude of the LOFAR spectrum of marine environment interference noise, that is the threshold $\varepsilon $ for determining whether the point $p_i$ is a breakpoint.

It can be analyzed from the cost function: when the point on the path passes or is close to the line spectrum, the sum of the point amplitude on the path increases, while the frequency gradient, the number of breakpoints and the cost function $O(\eta)$ decrease. The path which the target cost function is the smallest is considered to have a line spectrum, so the problem of line spectrum detection is transformed into the problem of finding the optimal path $\eta$ and minimizing the cost function about the path $\eta$.

\subsection{Sliding Window Line Spectrum Extraction Algorithm based on Multi-step Decision}

\begin{figure}[t]
  \centering
  \includegraphics[width=0.86\linewidth]{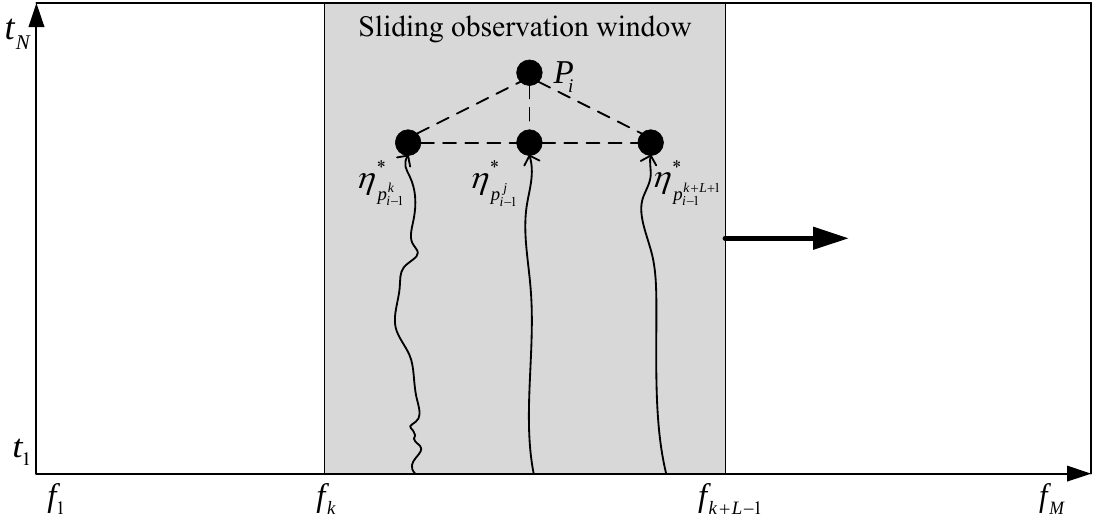}
  \caption{Frequency-domain sliding window multi-step decision dynamic tracking line spectrum.}
  \label{Fig:1}
\end{figure}

In this section, for the problem of minimizing the cost function mentioned in the previous section, a sliding window line spectrum extraction algorithm based on multi-step decision is used to search for the optimal path. As shown in Fig.~\ref{Fig:1}, in this algorithm, a window which can slide along the frequency axis and cover the whole time axis is set in the LOFAR spectrum. We search  the optimal path in this window. The reason for setting the window is that there may be multiple line spectrum co-existing in the LOFAR spectrum. By properly setting the size of the window, the search range of the path can be limited to a certain region of the LOFAR spectrum. Then the line spectrum in each window can be extracted, which can avoid that only the strongest spectral line is extracted in the whole LOFAR spectrum.



In order to cover a line spectrum in a search window, the size of the window is related to the following two points:

(1) The line spectrum width of the ships' radiated noise is related to its center frequency. The Doppler frequency shift caused by the ship's motion will also broaden the line spectrum to a certain extent, so the size of the window needs to ensure that the line spectrum is completely contained in the window;

(2) The size of the STFT frame, which comes from the process of calculating the LOFAR spectrum, determines the frequency resolution in the LOFAR. The size of the window can be calculated by combined with (1).


The specific steps of the sliding window line spectrum extraction algorithm based on multi-step decision are as follows:

(1)	Define the search window size $L$;

(2) Define the ternary vector of each point at time  $t_i$;

(3) From time $t_2$ to time $t_N$, find the optimal path from $2$ to $N$ line by line in the search window.

(4) Get the optimal path with length $N$ in the search window at time $t_N$:
\begin{equation}
O\left ( \eta ^{*} \right )=\underset{k\leq j\leq k+L-1}{min} O\left ( \eta _{P_{N}^{j}}^{*} \right )
.
\end{equation}

(5)	A counter is set for each time-frequency point in the LOFAR spectrum, and the counter value is initialized to 0. If the objective function value $O\left ( \eta ^{*} \right )$ corresponding to the optimal path $\eta ^{*}$ in the search window is greater than the threshold $\gamma$, the counter values corresponding to $N$ points on the optimal path will be respectively increased by $1$. The specific threshold value is:
\begin{equation}
	\gamma = \underset{1\leq r\leq M}{min} O\left ( \eta _{noise}^{r} \right )
	.
\end{equation}

(6)	Slide the search window with a step size of 1. Repeat the above steps until the observation window slides to the end position. The output count value graph is the line spectrum obtained by tracking.

The specific calculation steps are as follows:

(1) The length of the frequency axis in the LOFAR spectrums $M$. The start point is $f_1$, and the end point is $f_N$. The length of the time axis is $N$. The start point is $t_1$, and the end points $t_N$. The search window size is defined as $L$.


(2) Each point in the figure is defined as $P^j_i$, representing the time-frequency pixel on the $j$th column on the frequency axis and the $i$th row on the time axis, where $1 \le j \le M$, $1 \le i \le N$. $\eta _{P_i^j}^{*}$ represents the optimal path from $t_1$ to $t_N$ in the observation window, ${A(\eta _{P_i^j}^*),F(\eta _{P_i^j}^*),T(\eta _{P_i^j}^*)}$ defines as a set of ternary vectors for points $P_i^j$, and the triplet of each point at $t_1$ is initialized to $(a(P_1^j),0,0)$.

(3) From $t_2$ to $t_N$, find the optimal path with length from $2$ to $N$ in the search window line by line. In the figure, $P_i$ is set to any point in $t_1$, the start position of the observation window is $f_k$, and the corresponding end position is $f_{k+L-1}$. At $t_{i-1}$, the neighboring $L$ points of $P_i$ form a set as follows, $V({P_i}) = \{ P_{i - 1}^k, \cdots ,P_{i - 1}^{k + L - 1} \}$, the optimal path $\eta _{P_i}^{*}$ to the length $i$ of the point $P_{i}$ is obtained from the optimal path $\eta_{P_i^j}^{*}$ of $P_{i-1}^j \in V({P_i})$ , that is $\eta _{{P_i}}^* = \eta _{P_{i-1}^j}^* \cup \{{P_i}\}$, where $k \le j \le k + L - 1$ satisfies:
\begin{equation}\label{}
O(\eta _{P_i}^*) = \underset{P_{i - 1}^j \in V({P_i})}{min} O(\eta _{P_{i - 1}^j}^* \cup \{ {P_i} \})
,
\end{equation}
\begin{equation}\label{}
j = \underset{P_{i - 1}^j \in V({P_i})}{argmin} O(\eta _{P_{i - 1}^j}^* \cup \{ {{P_i}} \})
,
\end{equation}
where $\left \{ P_{i} \right \}$ represents the set of points $P_{i}$.

(4) At $t_N$, the optimal path of the $k$ points $P_N^j$ in the search window is $\eta_{P_N^j}^*$, where $k \le j \le k+L-1$, then the optimal path of length $N$ in the search window is:
\begin{equation}\label{}
O(\eta *) = \underset{k \le j \le k + L - 1}{min}O(\eta_{P_N^j}^*)
.
\end{equation}

(5) Set a counter for each time-frequency point in the LOFAR spectrum, and the counter value is initialized to 0. If the value of the objective function $O(\eta *)$ corresponding to the optimal path $\eta^{*}$ in the search window is greater than the threshold $\gamma$, we would consider that there is a line spectrum on the optimal path, and the counter values corresponding to the $N$ points on the optimal path are increased by $1$ respectively. The specific steps of threshold calculation are as follows:

First, the input of the algorithm is changed from the LOFAR spectrum of ship radiation noise to the LOFAR spectrum of marine environmental noise. The corresponding cost function $O\left ( \eta _{noise}^{r} \right )$ of the optimal path $\eta _{noise}^{r}$ in the $rth$ observation window is obtained, where $1\leq r\leq M-L+1$ then the threshold is:
\begin{equation}
\gamma = \underset{1\leq r\leq M}{min} O\left ( \eta _{noise}^{r} \right )
.
\end{equation}

(6) Slide the search window with a step size of $1$. Repeat the above steps until the observation window slides to the end. The output count value graph is the traced line spectrum.

\section{UNDERWATER TARGET RECOGNITION FRAMEWORK DESIGN}


\subsection{Design of underwater target recognition framework based on convolutional neural network }


Recently, CNN has proven its powerful capability in many fields, such as computer vision, nature language processing, and wireless physical layer \cite{liu2020location, yuan2020learning, liu2020deepresidual, xie2019activity, xie2020unsupervised, xie2020deep}.
Convolutional neural networks are deep feedforward neural networks that include operations such as convolution calculations, pooled sampling, and nonlinear activation \cite{goodfellow2016deep, hubel1962receptive}. Compared with the traditional feedforward neural networks like MLP, three strategies in CNN make use of the spatial correlation of data which include weight sharing, local receptive field and down sampling. They reduce the risk of over fitting, the defect of gradient disappearance the complexity and parameter size of the network. However, they improve the generalization ability of the network. CNN was first proposed by LeCun \cite{lecun1990handwritten} in 1990 and applied to the handwritten character detection system. In 2014, Szegedy \cite{szegedy2015going} proposed GoogleLeNet which introduced the inception module. Receptive fields of different sizes enhanced the adaptability of the network to scale. The improved version \cite{szegedy2016rethinking, szegedy2017inception} greatly reduces the parameter amount to enhance the nonlinearity of the network and speed up the calculation. The residual network was proposed by Kaiming. He \cite{he2016deep} in 2015 adopted the idea of Shortcut Connection (SC) to solve the problem of network degradation.

\begin{table}[]
	\centering
	\caption{The CNN NETWORK MODEL PARAMETERS OF MEASURED DATASET}
	\renewcommand\arraystretch{1.5}
	\begin{tabular}{|c|c|c|c|c|}
		\hline
		\multicolumn{5}{|c|}{Input layer (1024)*64*1}                                                                                                                                                                                                                                                                                                                                                                           \\ \hline
		\begin{tabular}[c]{@{}c@{}}Conv+\\ ReLU\\ (7*7)*16\\ stride=2*1\end{tabular} & \begin{tabular}[c]{@{}c@{}}Conv+\\ ReLU\\ (7*5)*16\\ stride=2*1\end{tabular} & \begin{tabular}[c]{@{}c@{}}Conv+\\ ReLU\\ (5*5)*16\\ stride=2*1\end{tabular} & \begin{tabular}[c]{@{}c@{}}Conv+\\ ReLU\\ (3*3)*16\\ stride=2*1\end{tabular} & \begin{tabular}[c]{@{}c@{}}MaxPool\\ (3*3)\\ Conv+ReLU\\ (1*1)*16\\ stride=1*1\end{tabular} \\ \hline
		\multicolumn{5}{|c|}{Fileter concatation}                                                                                                                                                                                                                                                                                                                                                                               \\ \hline
		\multicolumn{5}{|c|}{ReLU+MaxPool (3*3)}                                                                                                                                                                                                                                                                                                                                                                                \\ \hline
		\multicolumn{5}{|c|}{Conv+ReLU(5*5)*16 stride=2*1}                                                                                                                                                                                                                                                                                                                                                                      \\ \hline
		\multicolumn{5}{|c|}{MaxPool (3*3)}                                                                                                                                                                                                                                                                                                                                                                                     \\ \hline
		\multicolumn{5}{|c|}{Conv+ReLU(5*5)*16 stride=2*1}                                                                                                                                                                                                                                                                                                                                                                      \\ \hline
		\multicolumn{5}{|c|}{MaxPool(3*3)}                                                                                                                                                                                                                                                                                                                                                                                      \\ \hline
		\multicolumn{5}{|c|}{Conv+ReLU (3*3)*32 stride=2*2}                                                                                                                                                                                                                                                                                                                                                                     \\ \hline
		\multicolumn{5}{|c|}{MaxPool (3*3)}                                                                                                                                                                                                                                                                                                                                                                                     \\ \hline
		\multicolumn{5}{|c|}{Flattern}                                                                                                                                                                                                                                                                                                                                                                                          \\ \hline
		\multicolumn{5}{|c|}{Dense (4)}                                                                                                                                                                                                                                                                                                                                                                                         \\ \hline
	\end{tabular}
	\label{parameters_of_database}
\end{table}

From the LOFAR spectrum of the measured underwater acoustic signal which is extracted through multi-step judgment, we design a convolutional neural network structure according to its characteristics. The specific network parameters can be seen in Table~\ref{parameters_of_database}. For this CNN network structure, it refers some ideas of the Inception module which uses different sizes of convolution kernels and weighs the characteristics of the global and local information distribution. This network structure selects different convolution kernels and pooling kernels for preliminary feature extraction. The output of each sub-layer is cascaded and passes through several convolutional layers and pooling layers. Finally, the flatten layer flattens the feature map and the network completes the classification by the Dense layer. Convolution and pooling performed in parallel in the network obtain features of different information scales. The network has strong feature extraction capabilities for the positional relationship of line spectrum on different frequency points in the LOFAR spectrum.

The network parameters of CNN have been marked in the table. $(p*q)*r$ means the size of the convolution kernel is $(p*q)$, $r$ means the number of channels. $stride = m * n$ means the step size is $m * n$. Conv and MaxPool are convolution layer and max pooling layer respectively. CNN training and optimization hyperparameters are shown in Table~\ref{table_1}.

\begin{table}[]
	\centering
	\caption{CNN TRAINING, OPTIMIZATION HYPERPARAMETERS}
	\renewcommand\arraystretch{1.5}
	\begin{tabular}{|l|l|}
		\hline
		Optimizer         & adam                        \\ \hline
		Learning rate     & 0.01                        \\ \hline
		Number of samples & 200                         \\ \hline
		Training round    & 30                          \\ \hline
		Loss function     & Cross entropy loss function \\ \hline
	\end{tabular}
	\label{table_1}
\end{table}

\section{NUMERICAL RESULTS}

\subsection{Source of experimental data}

The experimental data used in this article is divided into two parts: The first part of the underwater acoustic database is named ShipsEar \cite{santos2016shipsear}, which was recorded by David et al. in the port of Vigo and it is vicinity on the Atlantic coast of northwestern Spain. The second part is based on the four types of signals simulated by the ship radiated noise. By mixing with the audio No. $81-92$ in the database which are treated as the pure marine environment background noise, the simulated actual ship radiated noise under different signal-to-noise ratios is obtained.

Vigo Port is one of the largest ports in the world with a considerable cargo and passengers. Taking advantage of the high traffic intensity of the port and the diversity of ships, it can record the radiated noise of many different types of ships on the dock, including fishing boats, ocean liners, Roll-on/Roll-off ships, tugboats, yachts, small sailboats, etc. The ShipsEar database contains $11$ ship types (marine environmental noise) and a total of $90$ audio recordings in "wav" format, with audio lengths varying from $10$ s to $11$ min.


\begin{table}[]
	\centering
	\caption{FOUR TYPES OF SHIP TARGETS}
	\renewcommand\arraystretch{1.5}
	\begin{tabular}{|l|l|}
		\hline
		W & Fishing boat, trawler, mussel harvester, tugboat, dredge \\ \hline
		X & Motorboat, pilot boat, sailboat                          \\ \hline
		Y & Passenger ferry                                          \\ \hline
		Z & Ocean liner, ro-ro ship                                  \\ \hline
	\end{tabular}
	\label{table_2}
\end{table}

By extracting and summarizing audios in the database, it is divided into four categories according to the size of the ship types collected which is shown in Table~\ref{table_2}. In addition, the date and weather conditions of the collected audios, the coordinates and driving status of the ship’s specific position, the number, depth and power gain of hydrophones, atmospheric and marine environmental data are also listed in detail. The information can be used as a reference in the study.


Because of military security considerations in the field of underwater target recognition, military databases are mostly kept secret. However, due to the inconvenience of collection and the high cost of civil databases, there are few public civilian databases for researchers to use. After the emergence of the ShipsEar database, it have been used in the application research of ship radiated noise separation, denoising, classification, etc. It is also common to use this database to complete research in the field of deep learning \cite{yang2018competitive, shen2018compression, yan2018resonance, ke2018underwater, chen2019new, yuan2019joint, ke2020integrated}.

\subsection{Experimental software and hardware platform}

The hardware platform and software support required to complete the deep learning experiment are shown in Table~\ref{table_3}.

\begin{table}[]
	\centering
	\caption{EXPERIMENTAL HARDWARE PLATFORM AND SOFTWARE SUPPORT}
	\renewcommand\arraystretch{1.5}
	\begin{tabular}{|c|c|}
		\hline
		lab environment                                                                                       & Configuration                                                                       \\ \hline
		operating system                                                                                      & Ubuntu 16.04                                                                        \\ \hline
		Graphics card                                                                                         & GTX 1080ti                                                                          \\ \hline
		\multirow{2}{*}{\begin{tabular}[c]{@{}c@{}}Programming language and\\ version\end{tabular}}           & Python 3.6                                                                          \\ \cline{2-2}
		& Matlab R2016b                                                                       \\ \hline
		\multirow{3}{*}{\begin{tabular}[c]{@{}c@{}}Deep learning library and\\ software toolbox\end{tabular}} & \begin{tabular}[c]{@{}c@{}}Keras 2.3\\ (tensorflow backed)\end{tabular}             \\ \cline{2-2}
		& \begin{tabular}[c]{@{}c@{}}Librosa Audio processing\\ library (python)\end{tabular} \\ \cline{2-2}
		& \begin{tabular}[c]{@{}c@{}}TQWT Toolbox\\ (Matlab language)\end{tabular}            \\ \hline
	\end{tabular}
	\label{table_3}
\end{table}

\subsection{Multi-step decision LOFAR line spectrum enhancement algorithm validity test}

In this section, the audio data of ShipsEar (a database of measured ship radiated noise) is used to verify the effectiveness of the algorithm.

For the signal decomposition algorithm based on resonance, the parameters setting for extracting high resonance components are ${Q_h} = 4$, ${r_h}= 3$, ${J_h} = 32$, and the parameters setting for extracting low resonance components are ${Q_l} = 1$, ${r_l}= 3$, ${J_l} = 3$.

\begin{figure}[t]
  \centering
  \includegraphics[width=0.86\linewidth]{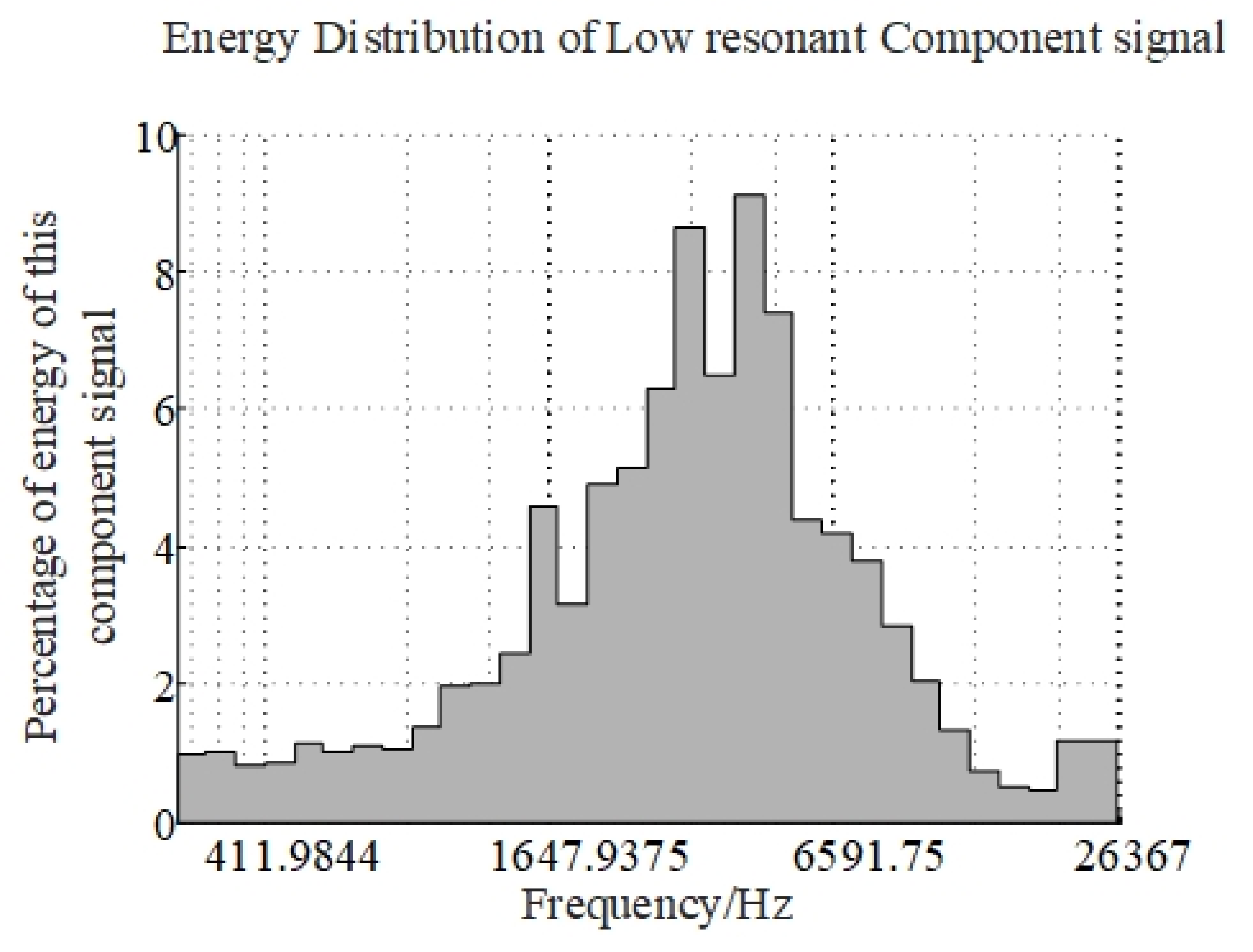}
  \caption{Percentage of total energy of each frequency band of low-resonance component signal.}
  \label{Fig:9}
\end{figure}

\begin{figure}[t]
  \centering
  \includegraphics[width=0.86\linewidth]{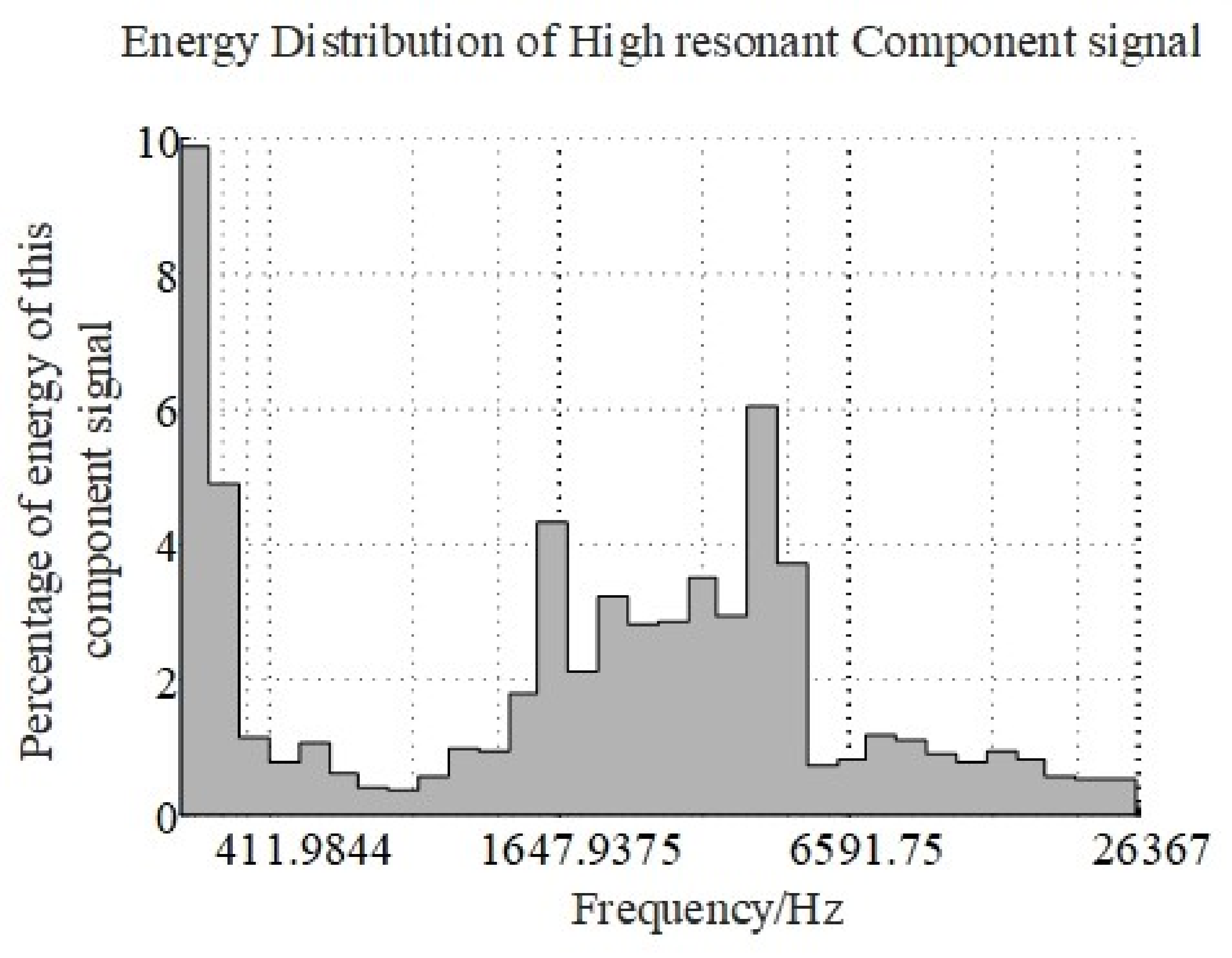}
  \caption{The percentage of total energy of each frequency band of the original signal.}
  \label{Fig:7}
\end{figure}

\begin{figure}[t]
  \centering
  \includegraphics[width=0.86\linewidth]{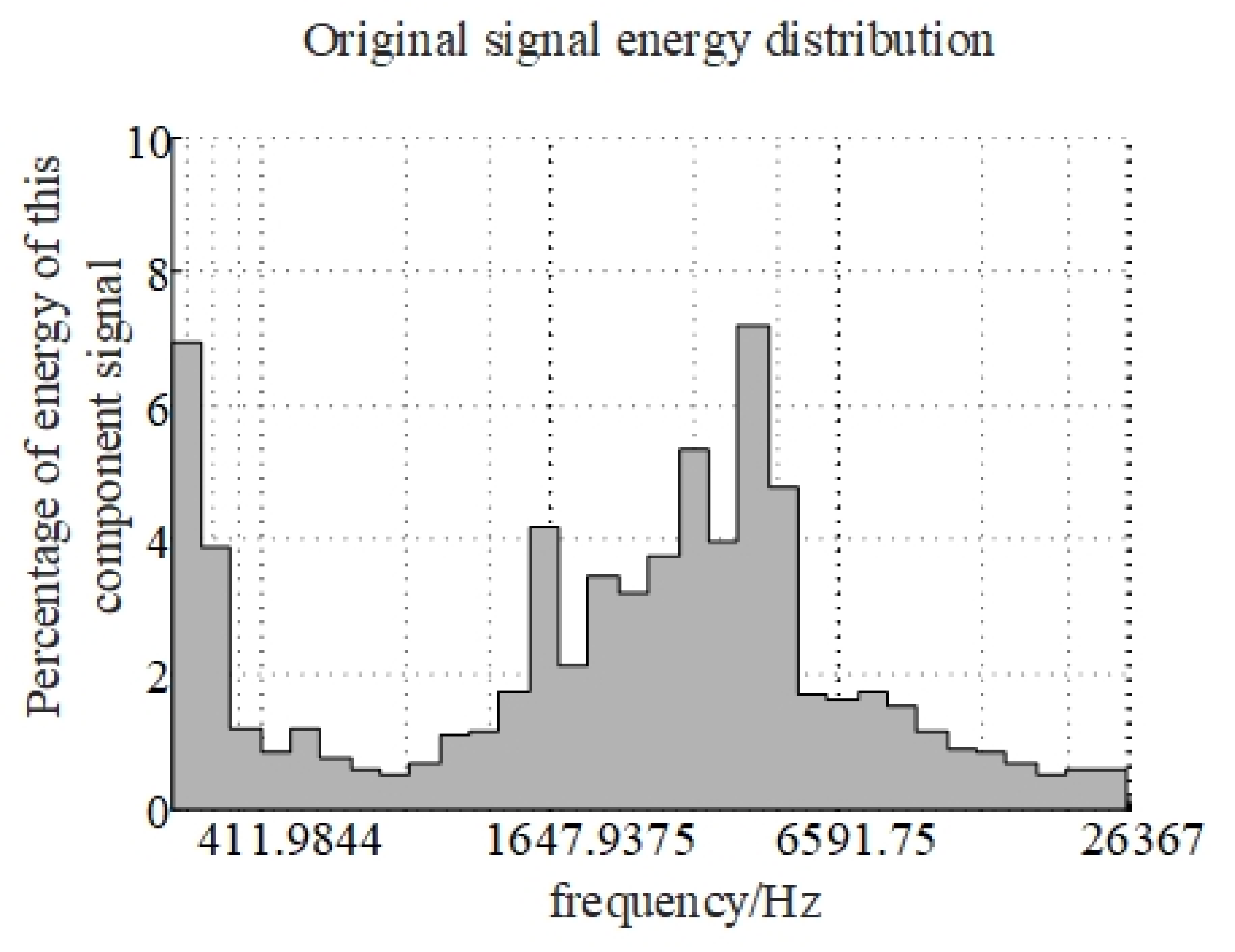}
  \caption{Percentage of total energy of each frequency band of high resonance component signal.}
  \label{Fig:8}
\end{figure}

From the energy percentage of each frequency band in Fig.~\ref{Fig:9}, the energy distribution of the low resonance component is mostly concentrated in the higher frequency band (greater than $1000$Hz), while the energy distribution in the low frequency band is very small. Comparing with Fig.~\ref{Fig:7}, we find that higher energy distribution of the original signal comes from the low resonance component. In Fig.~\ref{Fig:8}, most of the energy of the high resonance component is concentrated in the low frequency narrow band, and the narrow band energy distribution characteristic is usually regarded as a line spectrum. In previous studies, the low frequency line spectrum is the main manifestation of mechanical noise and propeller cavitation noise in the LOFAR spectrum. It is also an important basis for the identification of ship radiated noise. Therefore, the separated high resonance component retains the main features of underwater target recognition well.


In addition, Spectral Correlation Coefficient (SCC) \cite{hou1988spectrum} can also be used to measure the effectiveness of the RSSD algorithm. The physical significance of the spectral correlation coefficient is measuring the similarity of the power spectrum of the two signals, which is defined as follows:


\begin{equation}\label{}
{C_{A,B}} = \frac{{\int_{{f_1}}^{{f_2}} {{N_A}(f)} *{N_B}(f)df}}{{\sqrt {\int_{{f_1}}^{{f_2}} {N_A^2} (f)*\int_{{f_1}}^{{f_2}} {N_B^2(f)df} } }}
,
\end{equation}
where ${N_A}(f)$ and ${N_B}(f)$ represent the power spectrum of the two types of signals A and B, respectively. $f_1$ and $f_2$ represent the range of the power spectrum. This means that the radiated noise of the two types of ships with a higher degree of difference has a smaller spectral correlation coefficient. It can be seen from the Table~\ref{table_4} that the spectral correlation coefficients in the high-resonance components of signals A and B are smaller than their original spectral correlation coefficients. It means we can enhance the degree of difference between the two signals by extracting the high-resonance components of the signal.


\begin{table}[]
	\centering
	\caption{SPECTRAL CORRELATION COEFFICIENTS BETWEEN THE TWO TYPES OF ORIGINAL SIGNALS AND THEIR HIGH RESONANCE COMPONENTS}
	\renewcommand\arraystretch{1.5}
	\begin{tabular}{|c|c|c|}
		\hline
		signal & $S_{Original\_A}(t)$, $S_{Original\_B}(t)$      & $S_{high\_A}(t)$, $S_{high\_B}(t)$      \\ \hline
		$C_{A,B}$      & 0.7161 & 0.7074 \\ \hline
	\end{tabular}
	\label{table_4}
\end{table}

\begin{figure}[t]
	\centering
	\includegraphics[width=0.86\linewidth]{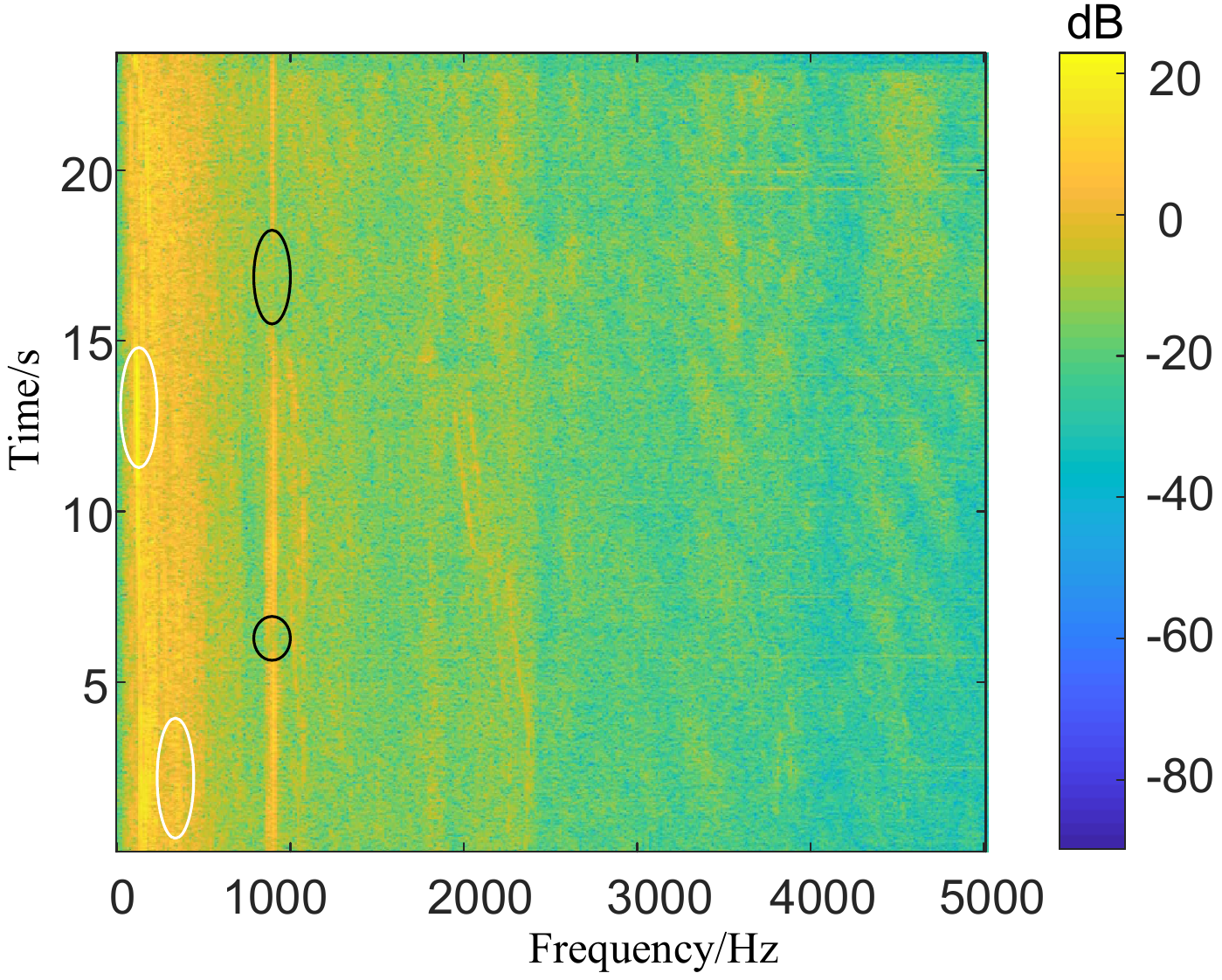}
	\caption{LOFAR spectrum of the original signal.}
	\label{Fig:10}
\end{figure}

\begin{figure}[t]
	\centering
	\includegraphics[width=0.86\linewidth]{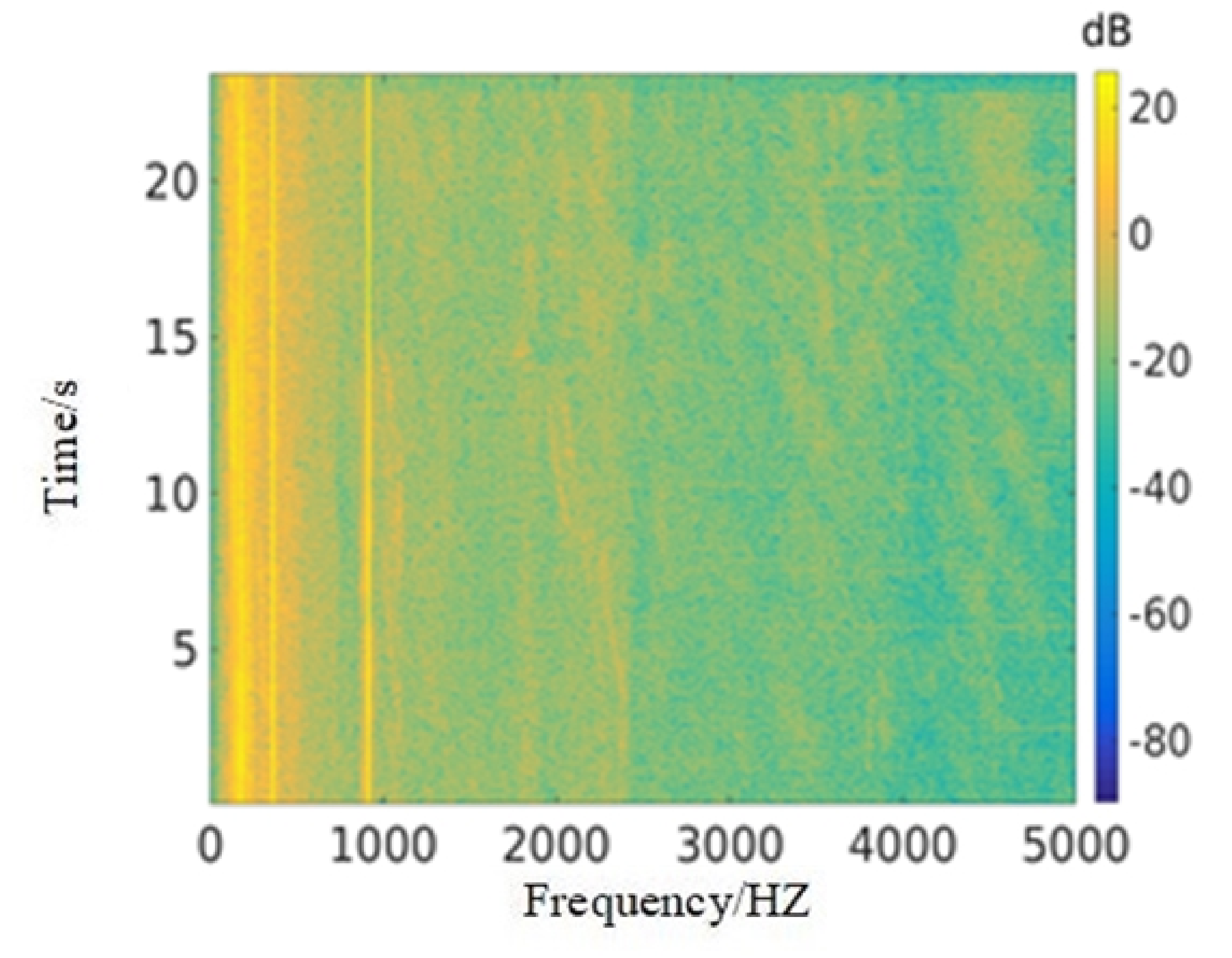}
	\caption{LOFAR spectrum after line spectrum enhancement.}
	\label{Fig:11}
\end{figure}

For the line spectrum enhancement algorithm based on multi-step decision, the experimental results are shown in Fig.~\ref{Fig:10} and Fig.~\ref{Fig:11}, which are the LOFAR spectrum of the original signal and the LOFAR spectrum after line spectrum enhancement. In Fig.~\ref{Fig:10}, there is a obvious line spectrum in the part marked by white circles, but the line spectrum is broken in the part marked by black circles. In Fig.~\ref{Fig:11}, the line spectrum indicated by the white circles are extended to completeness, and the vacant part of the line spectrum indicated by the black circles is also completed. Therefore, even if the line spectrum in the LOFAR spectrum has "breakpoints", "broken lines" or only a short line due to noise interference, the line spectrum enhancement algorithm can still extend and complete the line spectrum.


\subsection{Experimental verification of underwater target recognition based on convolutional neural network (CNN)}

a)~\textbf{CNN network offline training process}

According to the frame structure of underwater target recognition in Fig.~\ref{Fig:framework}, the specific settings and steps can be divided into:


1)~Read the high-resonance component signals in sequence, then windowing and framing the signal. We choose Hanning window (Hanning), and the window size is $2048$ (ie, FFT points are $2048$). The overlap between frames is $75\%$.


2)~The signal of each frame is normalized and decentralized. The power of the signal is uniform in time and the average value of the sample is $0$. It means the data is limited to a certain range, which can eliminate singular sample data. At the same time, it can also avoid the saturation of neurons and accelerate the convergence rate of the network.


\begin{table}[]
	\centering
	\renewcommand{\arraystretch}{1.5}
	\caption{VARIOUS SAMPLE TRAINING SETS AND TESTING SETS}
	\begin{tabular}{|c|c|c|c|}
		\hline
		\multirow{2}{*}{} & \multirow{2}{*}{ID}                                                               & Training set      & Testing set          \\ \cline{3-4}
		&                                                                                   & Number of samples & Number of samples \\ \hline
		W                 & \begin{tabular}[c]{@{}c@{}}46, 48, 66, \\ 73,  74, 75, \\ 80, 93, 94, \\ 95, 96\end{tabular} & 836               & 531               \\ \hline
		X                 & \begin{tabular}[c]{@{}c@{}}21, 26, 29, \\ 30, 50, 52, \\ 57, 70, 72, \\ 77, 79\end{tabular} & 837               & 516               \\ \hline
		Y                 & \begin{tabular}[c]{@{}c@{}}6, 7, 8, 10, \\ 11, 13, 14\end{tabular}                      & 1016              & 526               \\ \hline
		Z                 & 18, 19, 20                                                                          & 11449             & 603               \\ \hline
		Total             &                                                                                   & 3838              & 2176              \\ \hline
	\end{tabular}
	\label{table_5}
\end{table}

\begin{table}[]
	\centering
	\renewcommand{\arraystretch}{1.5}
	\caption{THE RECOGNITION ACCURACY RATE OF THE FOUR TYPES OF MEASURED SHIP RADIATED NOISE}
	\begin{tabular}{|c|c|c|c|c|c|}
		\hline
		Recognition rate & Class W & Class X & Class Y & Class  Z & average \\ \hline
		CNN              & 95.10\% & 87.60\% & 100.0\% & 97.78\%  & 95.22\% \\ \hline
	\end{tabular}
	\label{table_7}
\end{table}

3) First, perform Fourier transform on each frame signal. Second, take the logarithmic amplitude spectrum of the transformed spectrum and arrange it in the time domain. Last, take $64$ points on the time axis as a sample, which means getting a size of $1024 * 64$ LOFAR spectrum sample. The sampling frequency of audio is $52734$ Hz, and the duration of each sample is about $0.62$ s. The numbers of training and testing sets of various samples are shown in Table~\ref{table_5}. The ID in the table is the label of the audio in the ShipsEar database. The corresponding type of ship for each segment can be obtained according to the ID. The type of ship corresponding to audio is used as a label for supervised learning of deep neural networks.


4) The sample obtained in step (3) is treated with LOFAR spectrum enhancement as in the previous section. Then the LOFAR spectrum with enhanced line spectrum characteristics is obtained. The LOFAR spectrum is a two-dimensional matrix, which can be regarded as a single channel image. It will be used as the input layer of the convolutional neural network shown in the figure.



b)~\textbf{Identification of measured ship radiated noise}

The testing data set adopts the same feature preprocessing as the training set and inputs the trained model to complete the test.

\begin{figure}[t]
  \centering
  \includegraphics[width=0.86\linewidth]{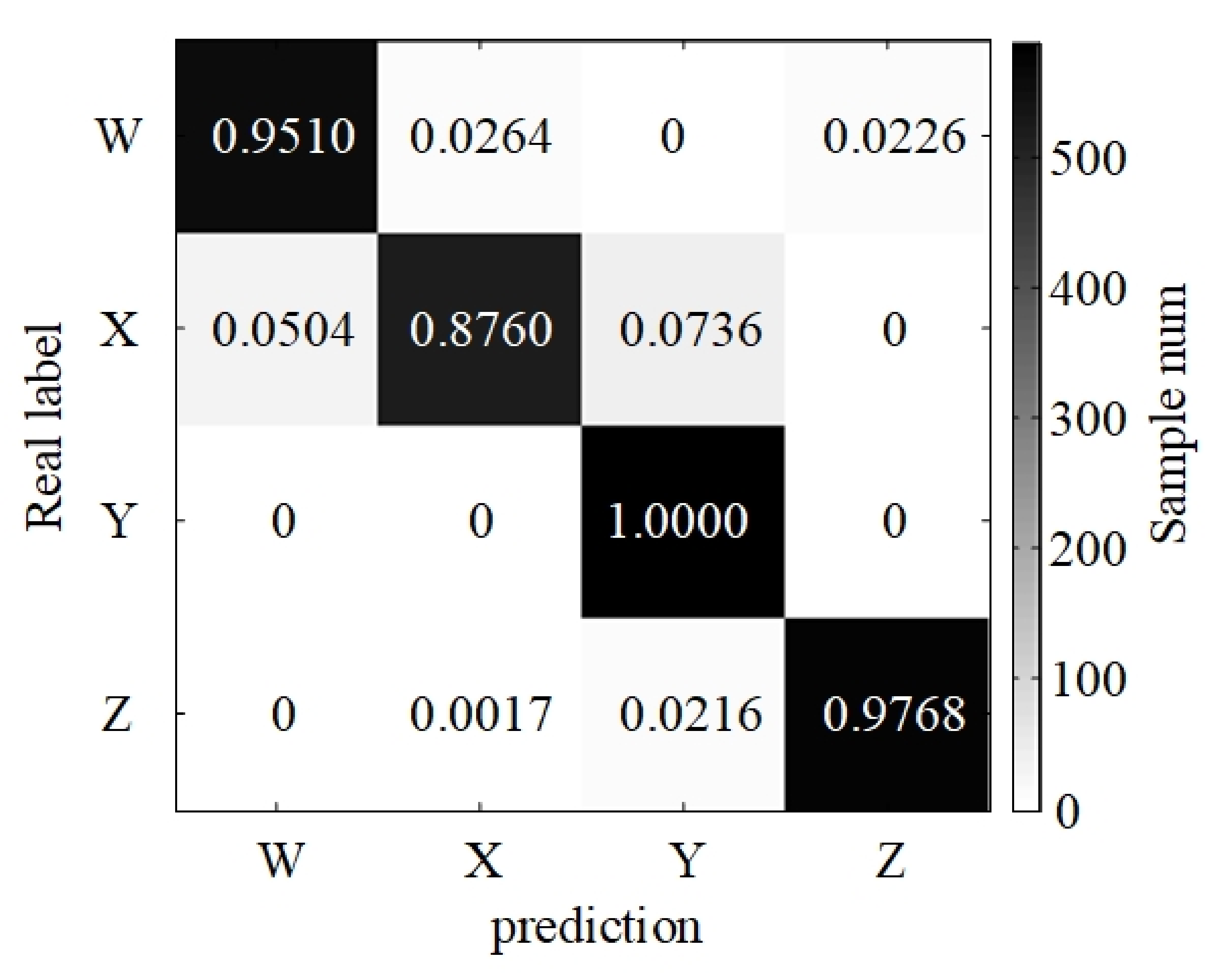}
  \caption{Confusion matrix of four types of measured ship radiated noise under CNN.}
  \label{Fig:12}
\end{figure}

Fig.~\ref{Fig:12} shows the standardized confusion matrix. The recognition accuracy of the radiated noise of the four types of ships is different. Among them, the recognition effect of the Y signal is the best and the recognition accuracy rate reaches $100.00\%$. The recognition accuracy rates of the W type and Z type are slightly worse and they are $95.10\%$ and $97.68\%$. Additionally, the recognition effect of the X type is the worst, which is only $87.60\%$. In summary, the total recognition accuracy rate is $95.22\%$. The recognition accuracy of four kinds of measured ship radiated noise is shown in Table~\ref{table_7}.

\begin{figure}[t]
  \centering
  \includegraphics[width=0.86\linewidth]{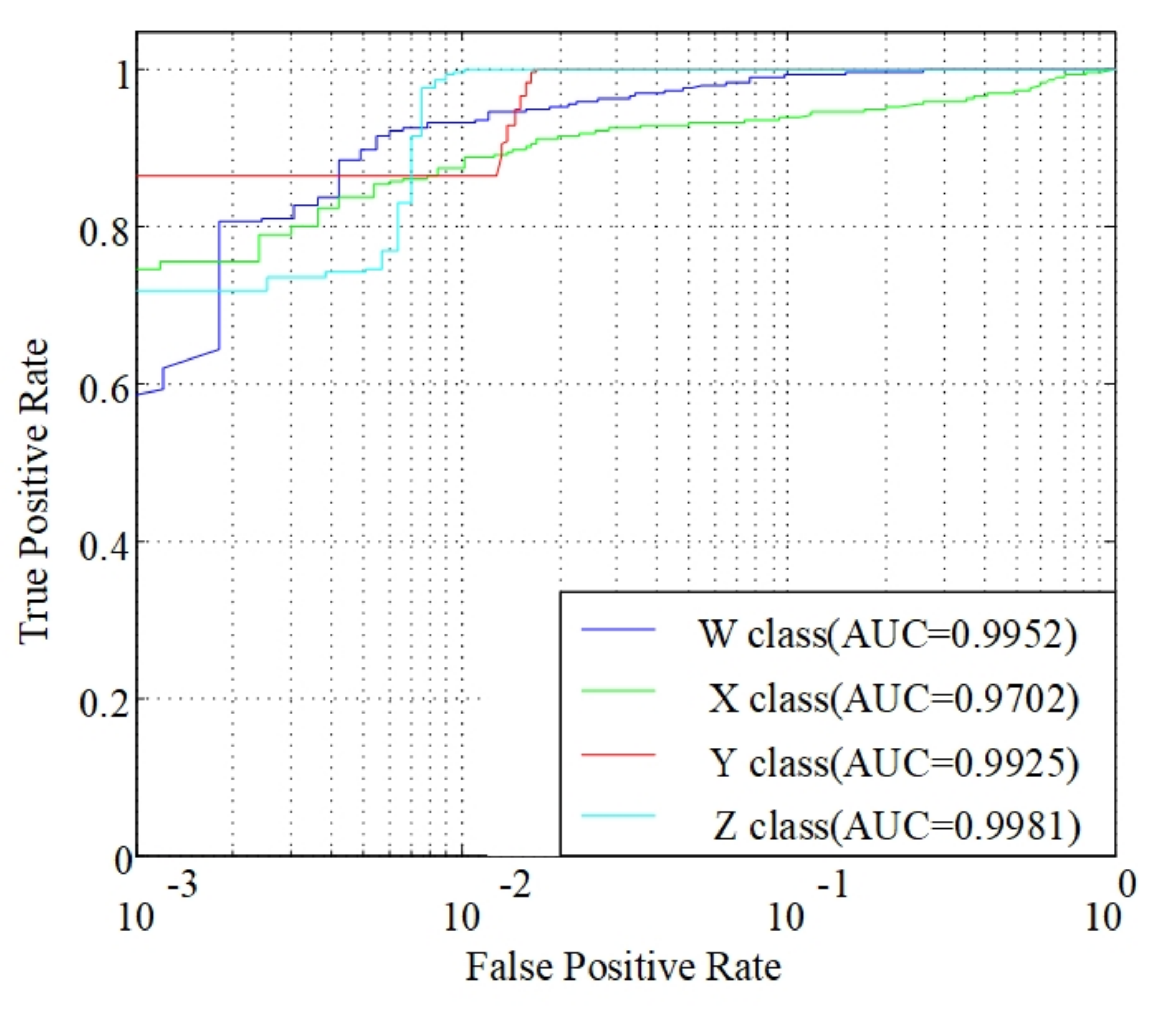}
  \caption{The ROC curve and AUC value of four types of measured ship radiated noise under CNN.}
  \label{Fig:13}
\end{figure}

Fig.~\ref{Fig:13} shows the ROC curve and the corresponding AUC value of the four types of signals. The horizontal axis uses a logarithmic scale to enlarge the ROC curve in the upper left corner. The ROC curves of the signals of W, Y, and Z are relatively close to the $(0, 1)$ point, and their classification effects are relatively good. However, the ROC curve of the signals of type X is closest to the $45$-degree line, so the classification effect is worst. Judging from the AUC value, the AUC of the Z-type signal is the highest, which reachs $0.9981$. The AUC values of the W-type and Y-type signals have respectively reached $0.9952$ and $0.9925$. The AUC value of the X-type signal is only $0.9702$. Therefore, the classification effect of the X-type signal is also inferior to the other three types of signals.


\section{CONCLUSION}

%

In this paper, we have studied underwater target recognition using the LOFAR spectrum. Firstly, a deep learning underwater target recognition framework based on multi-step decision LOFAR line spectrum enhancement is developed, in which we use CNN for offline training and online testing. Under the developed underwater target recognition framework, we then use the LOFAR spectrum as the input of CNN. Specially, on calculating the LOFAR spectrum of the high resonance component, we use the algorithm based on resonance and design the LOFAR spectrum line enhancement algorithm which is based on multi-step decision. To the best of our knowledge, the difference between the radiated noise of different types of ships is enhanced, and the broken line spectrum can be detected and enhanced. Finally, we conduct extensive experiments in terms of the detection performance, scalability, and complexity. The results have shown that the LOFAR-CNN method can achieve the highest recognition rate of $95.22\%$ with the measured ship radiation noise which can further improve the recognition accuracy compared with other traditional method.


\bibliographystyle{ieeetr}
\setlength{\baselineskip}{10pt}

\bibliography{Reference}

\end{document}